\documentclass[twocolumn]{aastex631}

\begin{document}

\title{
Rapid Expansion of the Young Type Ia Supernova Remnant 0519-69.0:\\ More Evidence for a Circumstellar Shell}

\correspondingauthor{Benson Guest}
\email{bguest1@umd.edu}

\author[0000-0003-4078-0251]{Benson T. Guest}
\affiliation{Department of Astronomy, University of Maryland, College Park, MD 20742, USA}
\affiliation{NASA Goddard Spaceflight Center, Greenbelt, MD 20771, USA}
\affiliation{Center for Research and Exploration in Space Science and Technology, NASA/GSFC, Greenbelt, MD 20771, USA}

\author[0000-0002-2614-1106]{Kazimierz J. Borkowski}
\affiliation{Department of Physics, North Carolina State University, Raleigh, NC 27695, USA}

\author[0000-0002-9886-0839]{Parviz Ghavamian}
\affiliation{ Department of Physics, Astronomy and Geosciences, Towson University, Towson, MD 21252, USA}

\author[0000-0003-3850-2041]{Robert Petre}
\affiliation{NASA Goddard Spaceflight Center, Greenbelt, MD 20771, USA}

\author[0000-0003-2251-6297]{Adrien Picquenot}
\affiliation{Department of Astronomy, University of Maryland, College Park, MD 20742, USA}
\affiliation{NASA Goddard Spaceflight Center, Greenbelt, MD 20771, USA}
\affiliation{Center for Research and Exploration in Space Science and Technology, NASA/GSFC, Greenbelt, MD 20771, USA}

\author[0000-0002-5365-5444]{Stephen P. Reynolds}
\affiliation{Department of Physics, North Carolina State University, Raleigh, NC 27695, USA}

\author[0000-0002-5044-2988]{Ivo R. Seitenzahl}
\affiliation{School of Science, University of New South Wales, Australian Defence Force Academy, Canberra, ACT 2600, Australia}

\author[0000-0003-2063-381X]{Brian J. Williams}
\affiliation{NASA Goddard Spaceflight Center, Greenbelt, MD 20771, USA}



\begin{abstract}
The nature of Type Ia supernovae remains controversial. The youngest remnants of Ia supernovae hold clues to the explosion and to the immediate surroundings. We present a third epoch of Chandra observations of the $\sim600$-year-old Type Ia remnant 0519-69.0 in the Large Magellanic Cloud, extending the time baseline to 21 years from the initial 2000 observations. We find rapid expansion of X-ray emitting material, with an average velocity of 4760 km s$^{-1}$. At the distance of the LMC this corresponds to an undecelerated age of 750 years, with the true age somewhat smaller. We also find that the bright ring of emission has expanded by 1.3\%, corresponding to a velocity of 1900 km s$^{-1}$ and an undecelerated age of 1600 years. The high velocity of the peripheral X-rays, contrasted with the modest expansion of the main X-ray shell, provides further evidence for a massive shell of circumstellar material.

\end{abstract}



\section{Introduction} \label{sec:intro}

Debate continues over the nature and progenitors of Type Ia supernovae. While observations of distant supernovae accumulate, the prompt emission at optical, UV, and IR wavelengths, and limits on radio and X-ray emission, sample only the very nearby supernova environment. As the supernova blast wave expands, it must eventually encounter any material which had been shed by the progenitor system at much earlier times, while differences in ejection velocities of ejecta from different layers have time to produce spatially resolvable separations, allowing explosion diagnostics (e.g. \cite{Hayato2010}). The youngest remnants show these interactions with minimal confusion from unrelated ambient material, offering the opportunity to test theories of Type Ia supernova origins.

The list of young Type Ia supernova remnants (SNRs) available for detailed study is disappointingly small: the Galaxy offers only the probable Ia remnant G1.9+0.3 \citep{Reynolds2008}, Kepler \citep{Reynolds2007}, Tycho \citep{Baade1945}, and SN 1006 \citep{Clark1977,Schaefer1996} from the last millennium. The Large Magellanic Cloud (LMC) exhibits three more: 0509-67.5, N103B, and the object of this study, SNR 0519-69.0 (henceforth 0519). Ages of all three are approximately known from light echos at $400 \pm 120$\,yr, $\sim 870$\,yr, and $600 \pm 200$\,yr, respectively \citep{Rest2005}. The well-known distance to the LMC of $\sim50$\,kpc \citep{Pietrzy2019} is much better determined than that to any of the Galactic remnants, while small enough to allow easily measurable proper motions of expansion. Here we present new X-ray observations of 0519 with Chandra, extending the time baseline for expansion measurements to 21 years.

SNR 0519--69.0 was discovered in the Large Magellanic Cloud (LMC) by an X-ray survey using the Einstein observatory \citep{Long1981}. Follow-up observations by \cite{Tuohy1982} confirmed the SNR nature and suggested a Type Ia origin. This was supported by X-ray observations with ASCA by \cite{Hughes1995}.

\cite{Borkowski2006} and \cite{Williams2011} used Spitzer data to infer that SNR 0519--69.0 has swept up several solar masses of material, while \cite{Seitenzahl2019} used models including the age and forward and reverse shocks as traced by broad [Fe \textsc{xiv}] emission to favour a Chandrasekhar-mass progenitor. Attempts to locate a surviving companion have been inconclusive \citep{Edwards2012}; however, \cite{Li2019} found a possible candidate.
The shock speed of SNR 0519-69.0 has been inferred from line widths observed with the Far Ultraviolet Spectroscopic Explorer (FUSE) by \cite{Ghavamian2007} to be $4000-6000$ km s$^{-1}$, and 2770 km s$^{-1}$ by \cite{Kosenko2010} using line broadening using RGS data from XMM-Newton. 
Proper motion studies in optical have been done using the Hubble Space Telescope by \cite{Hovey2018} and \cite{Williams2022}, the latter of which found two groupings of velocities in the measured filaments: a slower group with average velocity 1670 km s$^{-1}$ and a faster group with average velocity 5280 km s$^{-1}$. The difference was attributed to parts of the remnant interacting with material of a greater density than the average LMC ISM, likely circumstellar medium (CSM).

In this paper we present results from a proper motion study using a new epoch of Chandra observations from 2021. The baseline between observations is now 21 years from the earliest Chandra observations in 2000. We use the same methods as our recent paper on the LMC SNR 0509--67.5 \citep{Guest2022}.

\begin{figure}
    \centering
    \includegraphics[width=\columnwidth]{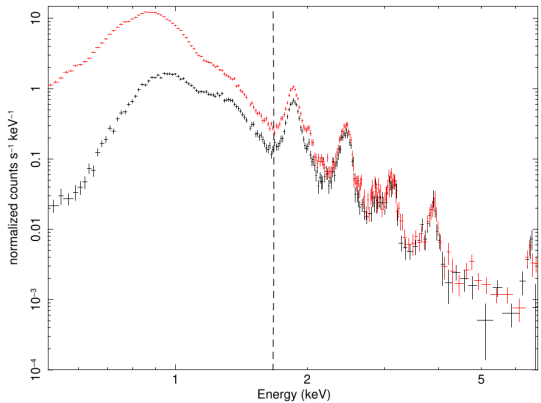}
    \caption{Spectrum taken from the 2000 observation (red) and the longest individual 2021 observation ObsID 24272 (black) showing the decrease in low energy sensitivity. The dotted line is at 1.7 keV.}
    \label{fig:FullRemnantSpectrum}
\end{figure}

\section{Observations}
A new epoch of Chandra observations (PI: B. Williams) was obtained over 15 segments from March 3 - November 2 2021, totalling 398 ks, and ranging in length from 11.1 to 36.4 ks. We select the arbitrary limit of 29.2ks for a useful observation segment (10 ks less than the earliest Chandra observation) and list the details in Table \ref{tab:Observations}. The initial Chandra observation (PI: S. Holt) consists of a single pointing on June 21 2000. In each observation SNR 0519-69.0 is placed on the S3 chip of the ACIS-S array near the optical axis of the telescope.

\begin{table}[]
    \centering
    \begin{tabular}{c c c}
	ObsID	&	Exp(ks)	&	Year		\\\hline
	118	&	39.2	&	2000		\\\hline
24263	&	29.7	&	2021		\\
24265	&	34.9	&	2021	\\
24272	&	36.4	&	2021	\\
24273	&	29.5	&	2021	\\
25061	&	29.7	&	2021	\\\hline\hline
    \end{tabular}
    \caption{\textit{Chandra} observations used in this work. }
    \label{tab:Observations}
\end{table}

\section{Image Alignment and Reprojection}
The standard procedure for aligning images relies on matching known point sources from external catalogues with point sources detected in an image, or aligning point sources detected in multiple images. Due to the position of SNR 0519--69.0 in the LMC, we did not find sufficient nearby point sources to perform this alignment. We instead return to the procedure used in our previous work \citep{Williams2018,Guest2022} and measure the diameter of the remnant.

\section{Measurements and Discussion}
\begin{figure}[t]
    \centering
    \includegraphics[width=\columnwidth]{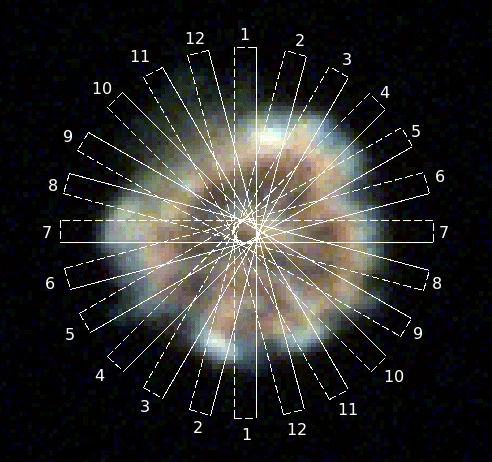}
    \caption{Chandra X-ray image of 0519--69.0, with 0.5-1.2 keV in red, 1.2-2 keV in green, and 2-7 keV in blue, overlaid with our 12 profile extraction regions as described in the text. Each region is 5 pixels wide where 1 pixel is the native Chandra pixel scale of 0$^{\prime\prime}$.492.}
    \label{fig:RGB-Image}
\end{figure}

The low-energy sensitivity of the Chandra X-ray Observatory has been significantly degraded by the buildup of contaminants on the optical blocking filter \citep{Marshall2004}. This difference becomes obvious when comparing the spectrum of the remnant taken at different epochs (Figure \ref{fig:FullRemnantSpectrum}).  We wish to compare datasets as
similar as possible in spectrum, so we restrict our energy range to higher energies where the responses are more similar.  While the response difference extends to $\sim 2$ keV, restricting observations to above 2 keV would eliminate the prominent Si line at 1.73 keV, which contains a substantial fraction of the total detected photon counts. We therefore filter the observations to 1.7 - 7 keV. We measured the diameter of the remnant along 12 projections which each cross the geometric center of the remnant and are spaced by 15 degrees (Figure \ref{fig:RGB-Image}). The measurements were made individually for each of the longest five 2021 observation segments (Table \ref{tab:Observations}) paired with the observation from 2000. These observations were chosen as they are within 10\,ks of the 2000 observation length, and we select the arbitrary limit of 29.2\,ks as the limit for a useful observation segment. 

\begin{figure}
    \centering
    \includegraphics[width=\columnwidth]{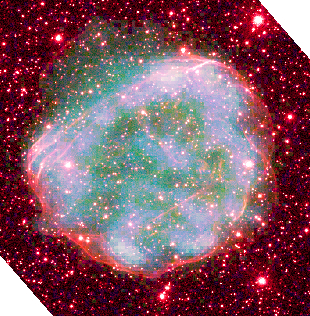}
    \caption{Multiwavelength image of SNR 0519-69.0. Green and blue show the merged Chandra image (0.5-1.95 keV and 1.95-7 keV, respectively); red shows the Hubble H$\alpha$ (F657N) image.}
    \label{fig:MultiWavelength}
\end{figure}

\begin{table*}[t]
    \centering
    \begin{tabular}{c c c c c}

Region	&	Avg shift (pixels)	&	Standard Deviation (pixels)	&	Velocity (km s$^{-1}$)	&	Standard Deviation (km s$^{-1}$)	\\\hline
1	&	1.84	&	-	&	5200	&	-	\\
2	&	1.69	&	0.14	&	4760	&	380	\\
3	&	2.53	&	0.53	&	7160	&	1510	\\
4	&	1.75	&	0.67	&	4940	&	1900	\\
5	&	1.07	&	0.32	&	3020	&	890	\\
6	&	1.53	&	1.21	&	4320	&	3430	\\
7	&	1.81	&	0.55	&	5290	&	1550	\\
8	&	1.5	&	0.51	&	3860	&	1440	\\
9	&	2.42	&	-	&	6830	&	-	\\
10	&	-	&	-	&	-	&	-	\\
11	&	-	&	-	&	-	&	-	\\
12	&	0.41	&	-	&	1170	&	-	\\\hline
\textbf{Average}	&	\textbf{1.7}	&	\textbf{0.44}	&	\textbf{4760}	&	\textbf{1300}	\\\hline\hline

    \end{tabular}
    \caption{Measurements of the expansion of the remnant along the 12 projection regions. Regions 10 and 11 do not have a measurement which is independent of small changes in the fitting window and have therefore been excluded. Regions 1, 9, and 12 have only one observation pair for which the measurement is independent of small changes in the fitting window. The velocity measurements have been included for completeness, but the average values only include the measurements for which multiple observation pairs yield values independent of small changes in the fitting window. The standard deviation for the average row is the deviations of the velocity measurements from regions 2-8.}
    \label{tab:Measurements}

\end{table*}

\begin{figure*}
    \centering
    \includegraphics[width=\textwidth]{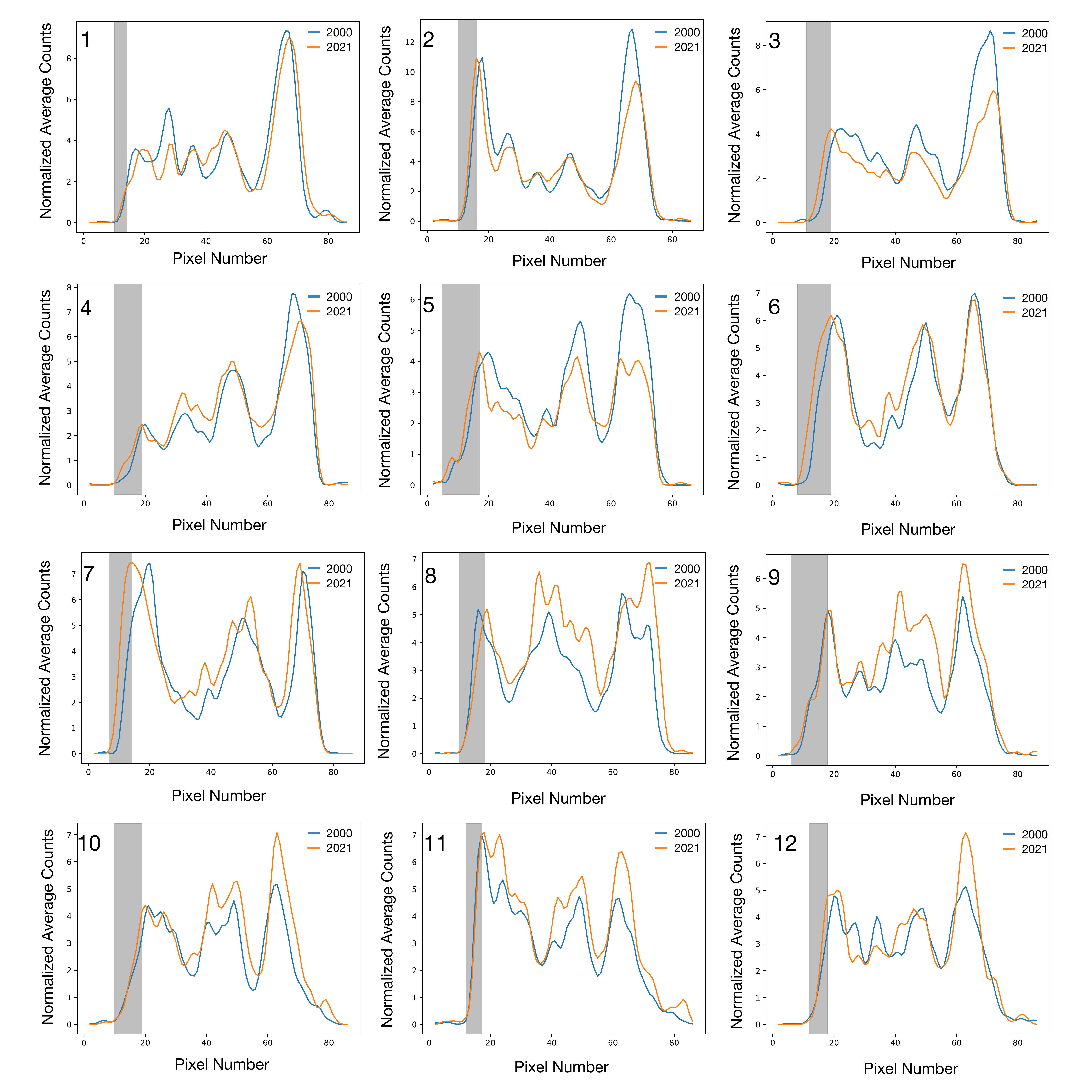}
    \caption{Sample projections from each region ordered clockwise starting with the north-south projection as numbered in Figure \ref{fig:RGB-Image}. The data are independently normalized prior to fitting the shift required for each side. The projections displayed are normalized for the left side and the shaded region in grey displays one of the chosen fitting windows (since each side is normalized independently, we show only the fitting window on the left side of the profile.)}
    \label{fig:SampleProjections}
\end{figure*}

The profiles of the X-ray counts distribution across a given region are normalized to align the top of the peak where the profile rises sharply from the background level. A window spanning the rise from the background level to the peak is selected by eye ranging from 4 - 10 pixels depending on the particular profile. The profiles are then shifted against each other on a grid of 0$^{\prime\prime}$.0049 corresponding to ~0.01 \textit{Chandra} pixels. We minimize the chi-squared value resulting from the difference between the two profiles over the defined window. The true window covering the rise from background to the peak of the shock-front is impossible to state with complete certainty, so we repeat the process 3 times for each measurement, changing the window boundaries by a pixel on either end each time. The best fit shift is then the average of each of the trial windows. As an additional check, the standard deviation of the best fit shifts for each observation pair measurement was calculated. If changing the fitting window led to a standard deviation greater than 0.1 pixels we did not include and carry this measurement forward in our average calculations since the measurement should be insensitive to the exact window choice to be considered reliable. The results are presented in Table \ref{tab:Measurements}.

Projections 10 and 11 do not have a measurement as small changes in the choice of fitting window had a significant effect on the results for every observation pair. These regions align with the northeast edge of the remnant (Figures \ref{fig:RGB-Image} \& \ref{fig:MultiWavelength}). The X-ray emission here does not show limb brightening. As a result, the emission is sufficiently faint that it does not yield a reliable profile from which to make measurements. This is most visible in the right side of profile 11 in Figure \ref{fig:SampleProjections}. Regions 1, 9, and 12 had only one pair of observations out of 5 which yielded a measurement which did not vary significantly based on the choice of window. The results from these measurements have been included in Table \ref{tab:Measurements} for completeness, but they have been excluded from the global average result. The standard deviation values listed in Table \ref{tab:Measurements} are the deviations of the average measurements from the individual observation pairs.

The average of the remaining projections yields an expansion velocity of $4760 \pm 1300$ km s$^{-1}$, where we have assumed a distance of 50 kpc. Including the measurements from regions 1, 9, and 12 has a minimal effect on the average, shrinking the global average to 4700 km s$^{-1}$. The faint northeast edge of the remnant in our projections 10 and 11 is likely due to low density and as a result may be hiding the fastest moving ejecta. Excluding a measurement from this region systematically lowers the measured global average. Measurements from \cite{Williams2022} in this region found velocities of $\sim4000-5000$ km s$^{-1}$, consistent with our global average such that we do not believe excluding these regions has greatly affected our result.

          

As an additional check, we attempted to calculate the average expansion of the remnant using contours generated for each epoch using the method from our previous papers \cite{Williams2016,Guest2022}. This is a completely independent method. To account for the differing response of the detector between epochs, we filter the energy to 2.3-7 keV. The contours are sensitive to the level of the background counts which depend on the observation length. We take this into consideration by selecting only counts detected during a portion of the 2000 exposure with the same number of counts as in the 2021 epoch. We drew a single contour around the entire remnant in ds9 using a contour level of 0.25 and a contour smoothing level of 4. The contours were converted to region files from which the area was calculated. We repeated this process for each of the 5 longest 2021 observations. We found that the contours did not encompass the faint emission in the north east edge of the remnant, and varied significantly between the 2021 observations in this area. Averaging our contour method results for the 5 longest 2021 observations compared with the 2000 epoch, we find the remnant has expanded with an average velocity of $2700 \pm 1800$  km/s. This is smaller than our more rigorous method above, however, because the contours do not extend to the edge of the emission in the northeast. The contours do not then fully encompass the entire SNR and as a result return a systematically smaller value for the expansion. The area of the longest 2021 observation segment contour equates to a circle with radius 15.3 arcseconds. Assuming free expansion with our average velocity from Table \ref{tab:Measurements}, we arrive at an upper limit on the age of 750 years, with the true age somewhat smaller since deceleration has almost certainly occurred.


\begin{table*}[t]
    \centering
    \begin{tabular}{l l c}
    \hline
        Remnant &  Velocity (km s$^{-1}$) & Age (yr)\\\hline
        0519-69.0 & 4760 & 750$*$\\
        0509-67.5$^{a}$ & 6120   & 620$*$ \\
        N103B$^{b}$ &  4170 & 850$*$ \\
        Tycho$^{c}$ & 2420 - 4800  & 450 \\
        Kepler$^{d}$ & 1700 - 7300  & 418 \\
        SN 1006$^{e}$ & 3000 - 5000 & 1016 \\\hline
        $*$ Free expansion \\
        $^{a}$ \cite{Guest2022} &  $^{c}$ \cite{Williams2016} & \\
        $^{b}$ \cite{Williams2018} & $^{d}$ \cite{Coffin2022} & \\
        $^{e}$ \cite{Katsuda2013,Winkler2014} & & \\\hline
    \end{tabular}
    \caption{Comparison of SNR 0519-69.0 with the X-ray proper motion expansion velocity measurements of other young Type Ia remnants. The ages inferred from the measured velocity assuming free expansion represent the upper limit on the age of SNRs 0519-69.0, 0509-67.5 and N103B, while the ages of Tycho, Kepler, and SN 1006 are from historical records \citep{Baade1943,Brown1952,Gardner1965}. The velocities listed for Tycho, Kepler, and SN 1006 are the range of velocities measured across the remnant. This is measurable for the nearby, large angular size Galactic remnants, not for the more distant LMC remnants for which an average velocity is given.}
    \label{tab:VelocityComp}
\end{table*}

Our high expansion velocity of X-ray-emitting material more closely resembles the higher H$\alpha$ shock velocities reported by \cite{Williams2022} ($\sim 5300$ km s$^{-1}$) than either the slower velocities from \cite{Williams2022} ($\sim 1700$ km s$^{-1}$) or the bulk ejecta velocities deduced from XMM-Newton RGS spectral line widths by \cite{Kosenko2010}.  They observe Gaussian line widths of $\sigma = 1873 $ km s$^{-1}$, of which some is due to thermal broadening.  \cite{Seitenzahl2019} found line widths (FWHM) of forbidden [Fe XIV] emission at 5303 \AA \,of $2460 \pm 100$ km s$^{-1}$ at a position at the edge of the observed [Fe \textsc{xiv}] MUSE image.  Since the radial velocity of iron at that position is expected to be small, the broad width ($\sigma = 0.43$(FWHM) $=$ 1045 km s$^{-1}$) indicates thermal broadening and allows us to extract the bulk velocity from the \cite{Kosenko2010} result (combining in quadrature), giving $\sigma({\rm bulk}) = 1550$ km s$^{-1}$.  Apparently the bright internal ejecta are expanding considerably more slowly than the periphery.

Now \cite{Williams2022} found no X-ray emission coincident with the fastest H$\alpha$ shocks, but our result shows that such fast velocities are true of X-rays as well.  The low ambient densities inferred by \cite{Williams2022} around most of the periphery suggest that the X-rays we see are the result of fast-moving ejecta rather than shocked circumstellar material.  Their velocity is unlikely to exceed the post-shock velocity behind the blast wave, or 3/4 of the forward-shock speed, roughly consistent with what we find.

However, the bright ring of emission, also ejecta, evidently expands much more slowly.  We used the method of \cite{Carlton2011} to measure its expansion between 2000 and 2021, arriving at 1.3 ($-0.3, +0.3$)\%\ (1$\sigma$ errors). This corresponds to a velocity of about 1900 km s$^{-1}$, much less than measured at the remnant's periphery, but in view of the rather large errors, consistent with the slower velocities found by \cite{Kosenko2010}, \cite{Seitenzahl2019}, and with the shell velocity of 1550 km s$^{-1}$ estimated from X-ray and optical line widths. The expansion rate of the bright shell, $0.063\%$ yr$^{-1}$, corresponds to a free expansion age of 1600 years, considerably larger than both the 750 year result derived from the periphery and the light echo derived age of 600 years \citep{Rest2005}.

Much of the evidence for slow ejecta in the interior of 0519 is based on lines of [Fe \textsc{xiv}], likely far inside the intermediate-mass elements (IMEs) such as Si and S which probably contribute most to the overall remnant brightness, given our energy acceptance range of 1.7 -- 7 keV (see Figure~\ref{fig:FullRemnantSpectrum}).  Significant velocity gradients in ejecta in Tycho's SNR (of comparable age to 0519 but less dynamically evolved) were reported from Suzaku data comparing line widths of Fe K$\alpha$ with those of IMEs Si, S, and Ar \citep{Hayato2010}: expansion velocities of the IMEs were found to be 17\% higher than those of Fe.  A qualitatively similar, but much larger, gradient might be occurring in 0519, with the outermost ejecta (here dominated by IMEs) expanding at several times the rate of inner Fe.

Thus there is substantial evidence for a considerably greater velocity of outermost ejecta than seen in the remnant's interior, and a departure from the simple 1D models of expansion into a uniform medium should be considered.  Evidently the remnant expansion has been nonuniform, with outer material less decelerated. A measure of the deceleration is the ratio $vt/r$ where v and r are the current velocity and radius, and t is the time since the explosion. Taking $t=600$yr from the light echo, we find $vt/r=0.4$ for the bright shell, while for the edge measurement we find $vt/r=0.8$. A remnant in the Sedov phase will have a ratio of 0.4, only slightly less than the 0.47 result of the hydrodynamic modeling of \cite{Seitenzahl2019}. The inner remnant results point to a dynamically advanced system, with nearly all ejecta shocked by now \citep{Seitenzahl2019}, while the edge measurements reveal a remarkable departure from uniform expansion, with the deceleration parameter twice as large there.

One possible explanation for this is a relatively recent, sudden expansion into lower-density surroundings, as would occur if the dense circumstellar medium inferred from optical evidence in \cite{Williams2022} were in the form of a shell, and parts of the forward shock have now broken through. Small ``blowout" regions visible in the H$\alpha$ image \citep{Williams2022} provide some evidence for this, along with the strong brightness variations with azimuth around the remnant edge. The model of \cite{Seitenzahl2019} yields a shock velocity of $v_S=2500$ km s$^{-1}$ such that we may estimate the post shock sound speed as $\sqrt{5}V_{S}/4 = 1400$ km s$^{-1}$ for $\gamma=5/3$. A rarefaction wave following the blast wave exit from a shell of CSM may accelerate material to several times the sound speed. The 4700 km s$^{-1}$ velocity we measure in the periphery minus the bulk expansion speed of 1900 km s$^{-1}$ relative to the sound speed yields $(4700-1900)/1400 = 2$, well within the limit accessible through rarefaction wave acceleration.
Further analysis of the complete 398 ks Chandra observation from 2021 should cast light on this interesting possibility, which will probably require multidimensional hydrodynamic modeling for fuller understanding. High spectral resolution observations of this remnant with {\it XRISM} or {\it Athena} would allow additional measurements of the kinematics from the line profiles, and observations with the James Webb Space Telescope (JWST) will trace the shocked CSM through infrared emission of dust grains. Finally, with continued monitoring over the coming decades at high-spatial resolution, the baseline for additional proper motion measurements will grow. We encourage future follow-up studies with {\it Chandra} or with the {\it Advanced X-ray Imaging Satellite (AXIS)}.


Comparing our value with X-ray proper motion velocities of other young remnants, we find that our measurement is in agreement with the fastest velocities measured in Tycho, SN 1006, and N103B. We list the velocity we measured along with measurements and ages from comparable Type Ia remnants in Table \ref{tab:VelocityComp}. We include the velocity ranges for the galactic remnants where the large angular size and suitable nearby point sources allowed for localized measurements. A future X-ray spectral study of SNR 0519-69.0 utilizing the total deep Chandra observation may draw connections with the circumstellar medium interaction seen in Kepler's SNR and N103B. As discussed in \cite{Williams2022}, SNR 0519-69.0 shows dense knots where the most prominent X-ray emission is coincident with regions of higher density as traced with Spitzer observations \citep{Borkowski2006}. Meanwhile, there is minimal X-ray emission from the fastest shocks seen in optical, namely the western edge of the remnant where the X-ray emission does not appear to extend out to the optical edge and the northeastern edge where the X-ray emission does not reach the outermost shock (Figure \ref{fig:MultiWavelength}).

\section{Conclusions}
We have observed SNR 0519--69.0 21 years after the initial Chandra observation in 2000. The lack of bright point sources in the field of view made the absolute astrometric alignment impossible, so we instead measured the change in the diameter of the remnant across many projections, yielding an average velocity of 4760 $\pm$ 1300 km s$^{-1}$. This corresponds to an undecelerated age of 750 years, with the true age somewhat lower. Continued monitoring of young SNRs in the Galaxy and the LMC is critical for understanding the early stage of their development, the stage most closely related to the explosion of the progenitor system. We encourage future observations of this remnant and others like it with \textit{JWST} in the infrared, {\it HST} and {\it LUVOIR} in optical bands and {\it Chandra}, {\it AXIS}, and {\it Lynx} at X-ray wavelengths.

\begin{acknowledgments}
We thank the referee for their careful reading of our paper. Their feedback improved the clarity of this work. B.G. acknowledges the material is based upon work supported by NASA under award number
80GSFC21M0002. P.G. acknowledges support from HST Grant GO-15989.005A to Towson University. K.B. and S.R. acknowledge support from Chandra grant GO1-22067A to N.C. State University.
\end{acknowledgments}

%

\vspace{5mm}
\facilities{CXO}





\bibliography{0519-Expansion}{}
\bibliographystyle{aasjournal}



\end{document}